\documentclass[prl,twocolumn,showpacs,superscriptaddress,preprintnumbers,amsmath,amssymb,nofootinbib]{revtex4-2}
\usepackage{bm}
\usepackage{dcolumn}
\usepackage{multirow}
\usepackage{ifpdf}
\usepackage{url}
\usepackage[utf8]{inputenc}
\usepackage{ulem}
\usepackage{xcolor}

\ifpdf
  \usepackage{graphicx}     
  \usepackage{hyperref}
  \usepackage{color}
\else     
  \usepackage[dvipdfmx]{graphicx}     
  \usepackage[dvipdfmx]{hyperref}
  \usepackage[dvipdfmx]{color}
\fi

\hypersetup{
colorlinks=true,
linkcolor=blue,
citecolor=blue,
urlcolor=blue
}

\begin{document}
\title{Probing the QCD phase transition with chiral mixing in dilepton production}
\author{Azumi Sakai}
\email{azumi-sakai@hiroshima-u.ac.jp}
\affiliation{
Physics Program, Hiroshima University,
Higashi-Hiroshima, Hiroshima 739-8526, Japan
}
\author{Masayasu Harada}
\email{harada@hken.phys.nagoya-u.ac.jp}
\affiliation{
Department of Physics, Nagoya University,
Nagoya, Aichi 464-8602, Japan
}
\affiliation{
Kobayashi-Maskawa Institute
for the Origin of Particles and the Universe, Nagoya University,
Nagoya, Aichi 464-8602, Japan
}
\affiliation{
Advanced Science Research Center, Japan Atomic Energy Agency,
Tokai, Ibaraki 319-1195, Japan
}
\author{Chiho Nonaka}
\email{nchiho@hiroshima-u.ac.jp}
\affiliation{
Physics Program, Hiroshima University,
Higashi-Hiroshima, Hiroshima 739-8526, Japan
}
\affiliation{
Department of Physics, Nagoya University,
Nagoya, Aichi 464-8602, Japan
}
\affiliation{
Kobayashi-Maskawa Institute
for the Origin of Particles and the Universe, Nagoya University,
Nagoya, Aichi 464-8602, Japan
}
\affiliation{
International Institute for Sustainability with Knotted Chiral Meta Matter (WPI-SKCM$^2$),
Hiroshima University,
Higashi-Hiroshima, Hiroshima 739-8526,
Japan
}
\author{Chihiro Sasaki}
\email{chihiro.sasaki@uwr.edu.pl}
\affiliation{
Institute of Theoretical Physics, University of Wroclaw,
plac Maksa Borna 9,
PL-50204 Wroclaw, Poland}
\affiliation{
International Institute for Sustainability with Knotted Chiral Meta Matter (WPI-SKCM$^2$),
Hiroshima University,
Higashi-Hiroshima, Hiroshima 739-8526,
Japan
}
\author{Kenta Shigaki}
\email{shigaki@hiroshima-u.ac.jp}
\affiliation{
Physics Program, Hiroshima University,
Higashi-Hiroshima, Hiroshima 739-8526, Japan
}
\affiliation{
International Institute for Sustainability with Knotted Chiral Meta Matter (WPI-SKCM$^2$),
Hiroshima University,
Higashi-Hiroshima, Hiroshima 739-8526,
Japan
}
\author{Satoshi Yano}
\email{syano@hiroshima-u.ac.jp}
\affiliation{
International Institute for Sustainability with Knotted Chiral Meta Matter (WPI-SKCM$^2$),
Hiroshima University,
Higashi-Hiroshima, Hiroshima 739-8526,
Japan
}

\date{\today}

\begin{abstract}
We perform a systematic study of dilepton emission in a hot QCD medium based on three different scenarios of chiral mixing, each of which yields a characteristic structure in the vector spectral function.
The in-medium spectral functions are accommodated into the state-of-the-art hydrodynamic simulations for a relativistic viscous fluid to calculate the dilepton production rate, fully accounting for the space-time evolution of a created fireball in relativistic heavy-ion collisions.
We demonstrate that the low-temperature theorem of chiral mixing extrapolated toward a chiral crossover, often used in the literature, leads to critical shortcomings: the inadequacy of width broadening, and a substantial overestimate of the dilepton yield maximized around the invariant mass of $M = 1.2$ GeV. The proper prescription offers a milder yet sizable increase in the window of $1.1 < M < 1.4$ GeV as the direct signature of chiral symmetry restoration.
\end{abstract}

\maketitle

{\it Introduction}
--
Spontaneous breaking of chiral symmetry in Quantum Chromodynamics (QCD) is centered on dynamical mass generation of hadrons composed of light valence quarks.
QCD matter created in relativistic heavy-ion collisions is a promising testing ground for exploring every change of hadron properties due to foreseen restoration of chiral symmetry 
driven by the QCD phase transition at high
temperature and/or net-baryon density~\cite{Hayano:2008vn,Rapp:2009yu,Fukushima:2013rx,Andronic:2017pug}.

Short-lived vector mesons are of particular importance as they are anticipated to carry imprints of chiral symmetry restoration (CSR) which would emerge in dilepton production rates.
In-medium modifications of the vector mesons are encoded in the spectral function, and the dilepton measurements in relativistic heavy-ion experiments indeed verified the existence of strong medium effects~\cite{CERES:2005,NA60:2006,STAR:2014,PHENIX:2016}, although it remains inconclusive that the observed modifications could be quantified as the straightforward signature of CSR.

The ideal signature is the mass degeneracy of chiral partners carrying opposite parities.
Especially favored is parity doubling of the $\rho$ meson and its counterpart, $a_{1}$ meson, because of their short lifetimes compared to the fireball.
However, the $a_{1}$ meson does not directly decay into a dilepton, so that its spectrum as a function of invariant mass is no longer possible to be reconstructed due to final state interactions.

A phenomenon called chiral mixing
between vector and axial-vector states
has hence attracted attention to overcome this experimental difficulty.
The chiral mixing effect is induced via a soft pion interacting with $\rho$ and $a_1$ states in a medium, and can be quantified in a model-independent way at low temperature and/or density~\cite{Dey:1990ba,Chanfray:1998hr,Kapusta:1993hq}.
Consequently, the in-medium $a_1$ can be indirectly accessed in the $\rho$ spectral function via the reactions of $a_{1}+\pi\rightarrow \rho$ and $a_{1}\rightarrow \rho+\pi$.
Dilepton spectra are thus supposed to carry the mixing effect and its thermal evolution toward CSR at which the $a_1$ meson is expected to be equally massive to the $\rho$ meson.

In the upcoming experiments at the CERN SPS~\cite{NA60:2022sze} and the LHC~\cite{ALICE:2022wwr},
the measurement of CSR via chiral mixing is set with a high priority.
In this letter, we shall address how different scenarios of CSR 
would possibly yield distinguishable results in the dilepton emission 
under the LHC condition.

There exist so far two major approaches to embed the chiral mixing into the in-medium spectral function.
One is to go beyond the low-temperature theorem toward a critical temperature of CSR~\cite{Harada:2008hj}, leading to the chiral mixing {\it resolved} at CSR based on model-independent aspects of chiral dynamics.
This is totally different from the naive expectation by extrapolating the mixing theorem to higher temperature, leading to
a {\it maximal} mixing~\cite{NA60:2018ckg,Geurts:2022xmk}. 
The other is to build up the axial-vector spectrum from the Weinberg sum rules with a reliable vector spectrum, e.g. the one compatible with the experiments~\cite{Hohler:2013eba}.
This approach requires some phenomenological assumptions, e.g. the energy dependence of the $a_1$ width as well as explicit inclusion of several low-lying states of vector mesons to saturate the sum rules.
It is notable that the two approaches arrive at a more or less the same conclusion - a single distinct maximum in the vector spectrum indicating strongly the chiral mixing resolved at CSR, in the striking contrast to the result based on the mixing theorem
causing inevitably two maxima.

Another element to accurately assess the degree of the phenomenon is
a reliable description of the space-time evolution of the medium created in relativistic heavy-ion collisions.
Bulk properties of the medium described by hydrodynamic models, especially temperature, volume and lifetime,
change dynamically in a short time scale,
$\sim$10 fm, and affect the dilepton yield.
Since the expected difference
of invariant mass spectra
is rather subtle with and without the meson mass modifications in any of the scenarios, quantitative
calculations anchored to firm field-theoretic bases are vital both in the QCD effective theories and hydrodynamic models.
Given the fact that the contemporary modeling of hydrodynamic evolution of a hot medium with temperature dependent transport coefficients is at hand~\cite{Okamoto:2017rup,Fujii:2022hxa},
we shall combine it with the hadronic spectral functions with chiral mixing that develops as a function of temperature,
to quantify the signatures of CSR relevant to the upcoming
relativistic heavy-ion experiments.

{\it Model}
--
The differential rate of dilepton emission at finite temperature $T$ from hadronic matter is related to the imaginary part
of the vector-current correlation function $G_V$ via
\begin{align}
\frac{dR_\mathrm{had}}{d^4q}(q;T) = \frac{\alpha_\mathrm{EM}^2}{\pi^3 M^2}
\frac{\mathrm{Im}G_V(q;T)}{e^{q_0/T}-1}\,,
\end{align}
where $\alpha_\mathrm{EM} = e^2/4\pi$ represents the electromagnetic coupling constant and
$M=\sqrt{q_0^2-|\bm{q}|^2}$ the invariant mass with energy $q_0$ and three-momentum $\bm{q}$ of
a virtual photon.
We shall construct the main input, the spectral function $\mathrm{Im}G_V$, based on a prescription guiding the chiral mixing
to the proper destination from low $T$ towards the chiral crossover $T_\chi$~\cite{Harada:2008hj}.
The interaction vertex of the $\rho$, $a_1$ and
$\pi$ mesons,
responsible for the chiral mixing in a medium, is uniquely represented in terms of
the mass difference, $\delta m = m_{a_1} - m_\rho$, or equivalently the pion decay constant $f_\pi$ 
that is the order parameter of chiral symmetry breaking and its restoration.
One readily finds that the chiral mixing vanishes at $T_\chi$ because of the degenerate $\rho$ and $a_1$ states.
In the following calculations, we will utilize a $T$-dependent $\delta m$ which is vanishing toward $T_\chi$ with the mean-field critical exponent,
as proposed in Ref.~\cite{Harada:2008hj}, whereas the $\rho$ meson mass is kept constant throughout this paper.
We note that the explicit breaking does not lead to a particular significance. 
Finite pion mass $m_\pi$ results in a slight mass difference of $\delta m = 3$ MeV at $T_\chi$, and the residual chiral mixing is negligible~\cite{Harada:2008hj}.

To differentiate the signature of CSR
from the bulk, we consider three scenarios:
(1) The case without CSR:
Thermal corrections to the spectral function are encoded via meson loops, while $\delta m$ is kept constant at any $T$.
(2) The case with CSR:
The in-medium spectral function is calculated with the thermal $\delta m$, leading to a gradual disappearance of the chiral mixing at higher $T$.
(3) The case with non-degenerate $\rho$ and $a_1$ (hereafter ``false CSR''):
This relies crucially on the validity of the low-temperature theorem near $T_\chi$.
The model-independent formulae for $G_V$ and
its parity counterpart
$G_A$
are given at low $T \ll m_\pi$ by
\begin{align}
G_V(q;T) &= \left( 1 - \epsilon \right) G_V(q;0) + \epsilon \,G_A(q;0) \ ,
\nonumber\\
G_A(q;T) &= \epsilon \, G_V(q;0) + \left( 1 - \epsilon \right) G_A(q;0) \ ,
\label{eq:theorem}
\end{align}
with $\epsilon= T^2/(6 f_\pi^2)$ being the mixing parameter.
When the $\epsilon$ could reach the maximal value of $1/2$, $G_V = G_A$ would be realized at $T = \sqrt{3}\,f_\pi = 160$ MeV.
The corresponding spectral functions would become identical as well, but this is a {\it fake} restoration. The reason is rather obvious:
The restored symmetry requires the vanishing chiral mixing as explained above, and this ensures simultaneously the degenerate $\rho$ and $a_1$,
the decay channel of $a_1 \to \rho\pi$ closed and the degenerate spectral functions in the vector and axial-vector sectors.
The maximal mixing is clearly incompatible with the degenerate $\rho$ and $a_1$ driven by a drastic change of QCD ground state emerging at the chiral phase transition.
Thus,
it fails to capture the non-trivial physics via a naive extrapolation of the low-temperature theorem.
It is odd to occur, but based on the scenario with false CSR it is often argued that the chiral mixing generates an increase
of the dilepton yield by $20$--$30$\% in
the invariant mass between $1$ and $1.4$~GeV as a signature of CSR~\cite{NA60:2018ckg,Geurts:2022xmk}.

Here we shall demonstrate how modifications due to the CSR emerge in the spectral function near $T_\chi$, and quantify a relative difference
among the three scenarios.
\begin{figure}[htb]
\includegraphics[width=0.45\textwidth]{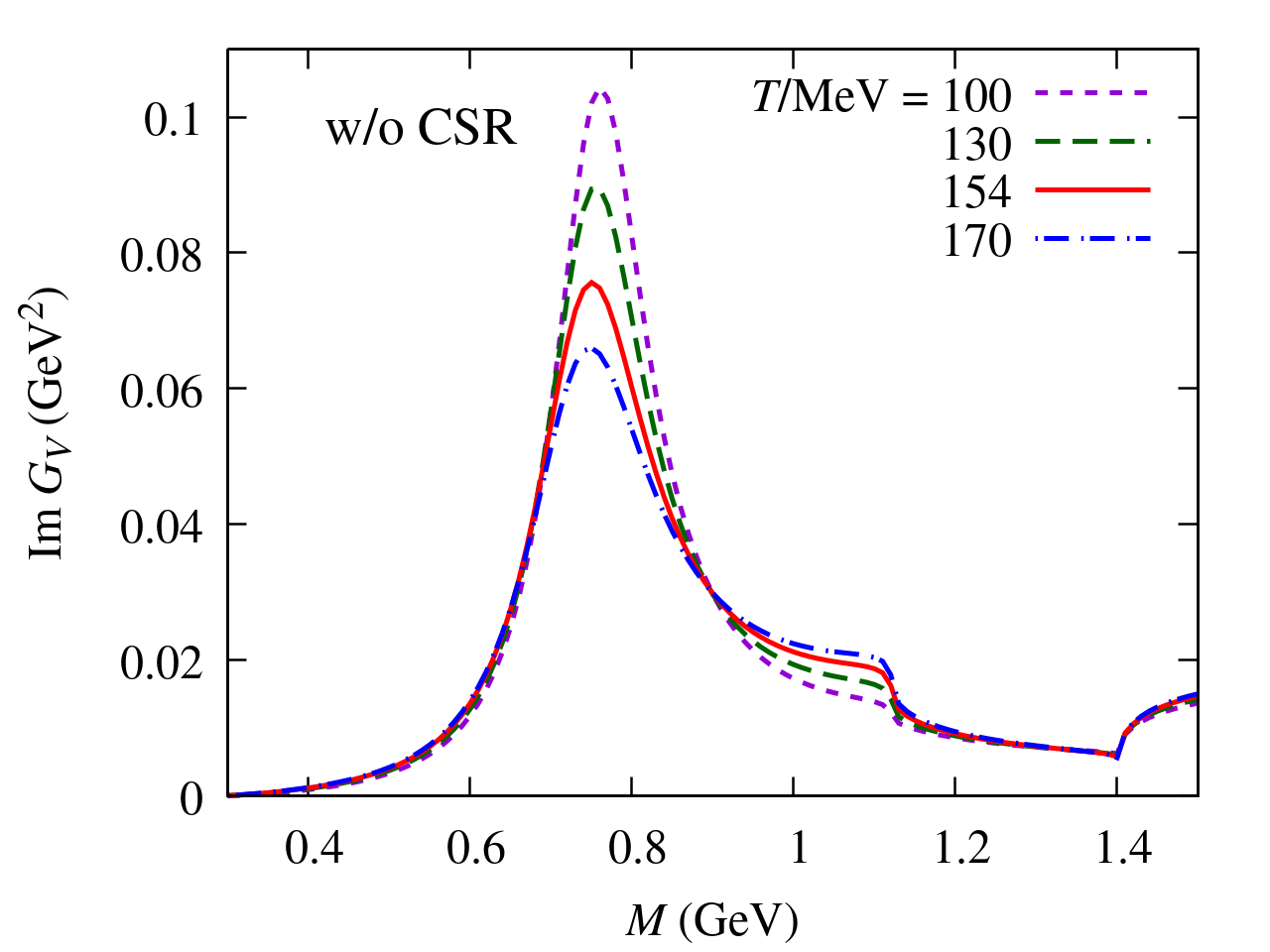}
\caption{
(Color online). The temperature dependence of
vector spectral function without chiral symmetry restoration.
We improved the calculations done in Ref.~\cite{Harada:2008hj} taking the updated value of $T_\chi = 154$~MeV
determined in lattice QCD~\cite{Aoki:2009sc,Bazavov:2011nk}.
}
\label{Fig:spectra_woCSR}
\end{figure}
%
Figure \ref{Fig:spectra_woCSR} shows
the vector spectral function at various temperatures without CSR but the chiral mixing fully implemented.
The position of the $\rho$ meson peak receives a slight shift due to the thermal effect produced mainly via pion loops,
whereas its strength is more reduced at higher $T$.
The chiral mixing enters the vector spectrum via the $a_1$-$\pi$ loop diagrams, which results in the threshold effects above $M \simeq 1$~GeV:
A shoulder at $M = m_{a_1} - m_\pi = 1.1$~GeV and a bump developing above $M = m_{a_1} + m_\pi = 1.4$~GeV arise
due to two kinematically allowed processes, $\rho + \pi \to a_1$ and $\rho \to a_1 + \pi$, respectively~\cite{Harada:2008hj}.
Although the shape of the spectral function is somewhat modified, the locations of the $\rho$ peak and the onset of threshold effects
are unmodified. This is because the bare masses of $\rho$ and $a_1$ states are set to be constant at any $T$, comprising the scenario
without CSR as our baseline.

\begin{figure}[htb]
\includegraphics[width=0.45\textwidth]{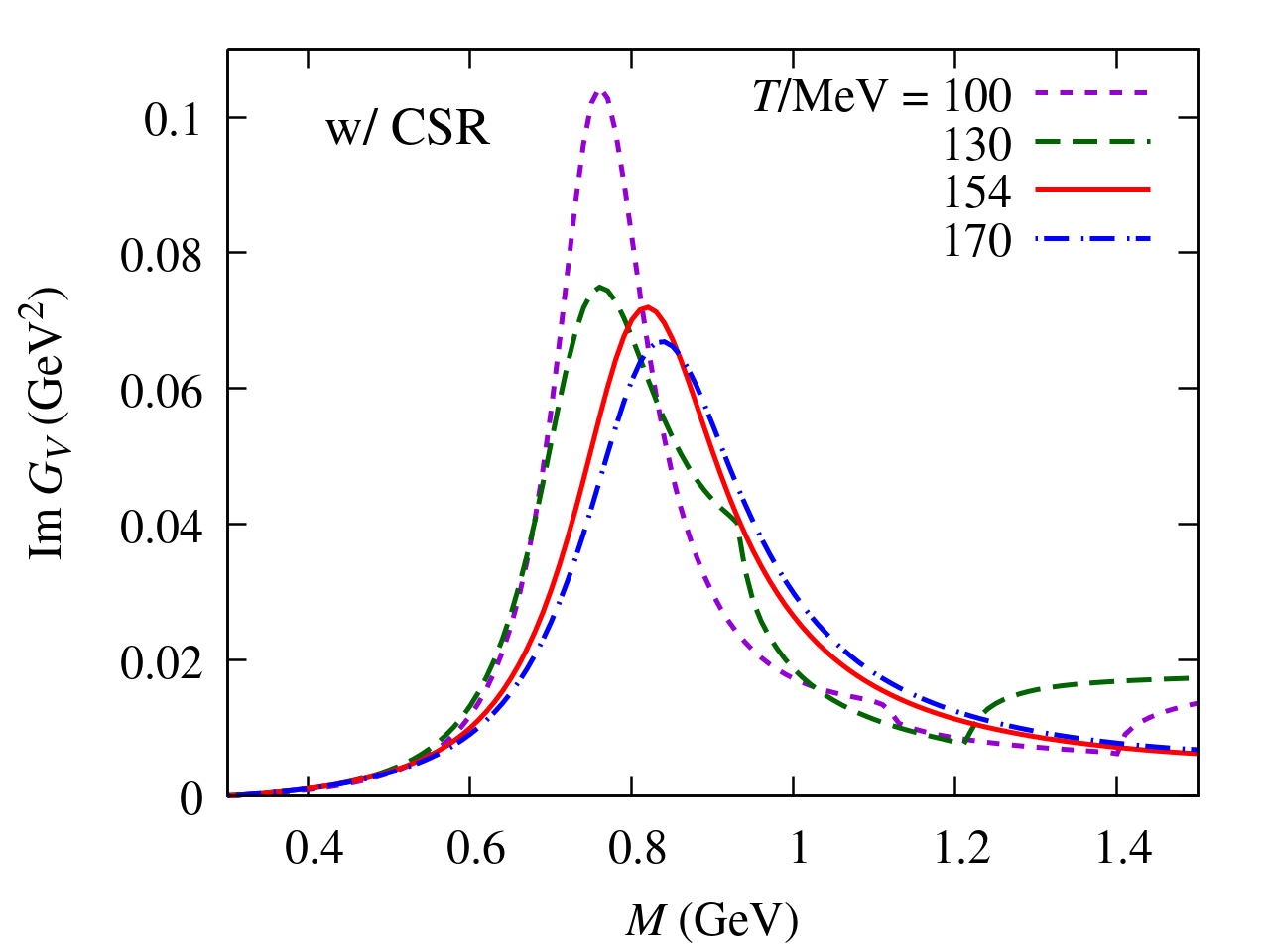}
\caption{
(Color online).
The same as in Fig.~\ref{Fig:spectra_woCSR}, but with chiral symmetry restoration.
}
\label{Fig:spectra_wCSR}
\end{figure}
The proper implementation of CSR modifies the vector spectrum substantially.
One finds in Fig.~\ref{Fig:spectra_wCSR} that the $\rho$ peak is shifted upward and the onset of threshold effects downward at higher $T$.
The former is induced by the pion-loop diagram carrying the coefficient of $(1 + m_\rho^2/m_{a_1}^2)$ which changes its strength
as $m_{a_1}$ approaches $m_\rho$.
The latter appears also via the same physics, i.e. the $a_1$ meson becomes lighter.
Those effects are eventually merged at $T_\chi$ to form a single maximum in the spectrum, clearly indicating the degenerate $\rho$ and $a_1$ states.

\begin{figure}[htb]
\includegraphics[width=0.45\textwidth]{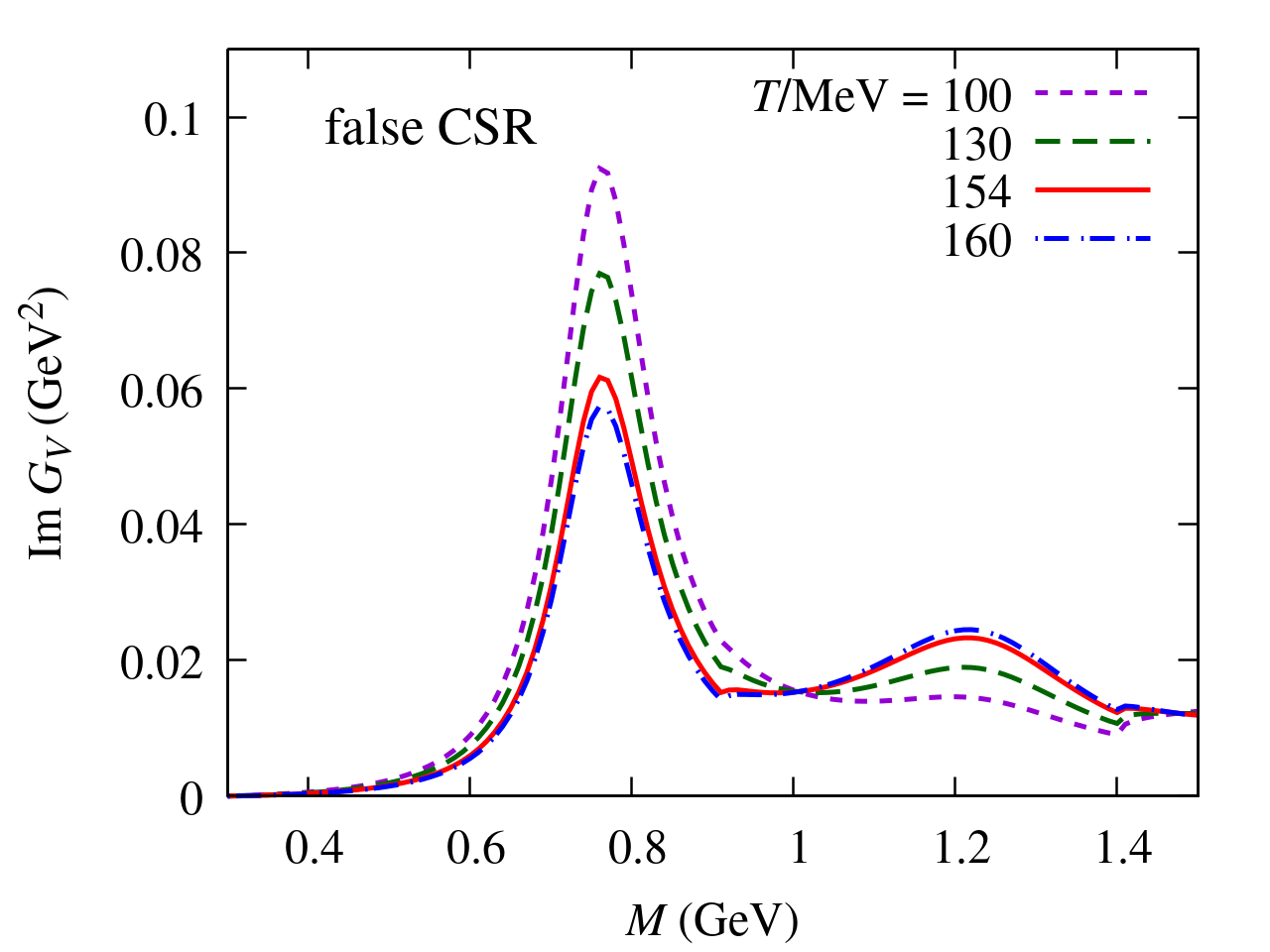}
\caption{
(Color online).
The same as in Fig.~\ref{Fig:spectra_woCSR}, but with false chiral symmetry restoration.
The mixing parameter $\epsilon$ reaches $1/2$ at $T=160$~MeV.
}
\label{Fig:spectra_DEI}
\end{figure}
Finally, the spectral function with
the false CSR calculated from Eq.~(\ref{eq:theorem}) is presented in Fig.~\ref{Fig:spectra_DEI}.
Since the entire thermal effects enter the current correlator only via the mixing parameter $\epsilon$, a peak on the left for the $\rho$ and a bump
on the right for the $a_1$ do no change their pole positions, but do their strengths according to the mixing parameter $\epsilon$ at a given $T$.
The spectral function in this scenario exhibits a striking contrast to the case with CSR shown in Fig.~\ref{Fig:spectra_wCSR}.
There exists a clear overestimate at around $M = 1.2$~GeV, lying in the primary window of $1 < M < 1.4$~GeV where the totally distinct structure
is found in Fig.~\ref{Fig:spectra_wCSR}.


The dilepton production rate
is integrated
over the whole space-time evolution
of the medium created
after the collision.
To describe a smooth crossover between the quark-gluon plasma (QGP) and hadronic phases at vanishing net-baryon density,
we utilize a combined production rate proposed in Ref.~\cite{Monnai:2019vup}:
\begin{align}
\nonumber
\frac{dR}{d^4 q} &=
\frac{1}{2}\left(1 - \tanh\frac{T - T_\chi}{\Delta T}\right)
\frac{dR_\mathrm{had}}{d^4 q}\\
&+\frac{1}{2}\left(1 + \tanh\frac{T - T_\chi}{\Delta T}\right)
\frac{dR_\mathrm{QGP}}{d^4 q}\,,
\label{Eq:total_rate}
\end{align}
with $\Delta T = 0.1T_\chi$.
The dilepton emission rate from the QGP medium due to $\bar{q}q$ annihilation is given in the Born approximation by
\begin{align}
\frac{d^4 R_\mathrm{QGP}}{d^4 q}(q;T) &= \frac{\alpha_\mathrm{EM}^2}{6 \pi^4}
\frac{1}{e^{q^0/T}-1}
\left\{1-\frac{2T}{|\bm{q}|}\ln\left[\frac{n_-}{n_+}\right]\right\},\\
n_\pm &= 1 + \exp\left[-\frac{q^0\pm|\bm{q}|}{2T}\right]\,.
\end{align}

{\it Results}
--
We use the state-of-the-art model of relativistic viscous hydrodynamics to deal with a dynamic evolution of QCD matter~\cite{Okamoto:2017rup}
in which the bulk and shear viscosities depending on temperature are taken into account.
A parametric initial condition model, TRENTo, is employed to obtain the profiles of initial entropy density~\cite{Moreland:2014oya,Ke:2016jrd}
with regulating the normalization parameter $N$ and the interpolating parameter $p$ between different initial conditions to reproduce
the LHC data~\cite{Fujii:2022hxa}.
Equations of state in lattice QCD simulations, parameterized in the hadronic and QGP phases~\cite{Bluhm:2013yga}, are utilized.
The hydrodynamic expansion starts at an initial time of $\tau_0=0.6$~fm under the assumption that
thermal equilibrium is achieved by then.
Our hydrodynamic calculations are carried out by generating $20$ events for Pb+Pb collisions with the collision energy
$\sqrt{s_\mathrm{NN}} = 2.76~\text{TeV}$
in the centrality window of 0--5\%.

\begin{figure}[htb]
\includegraphics[width=0.45\textwidth]{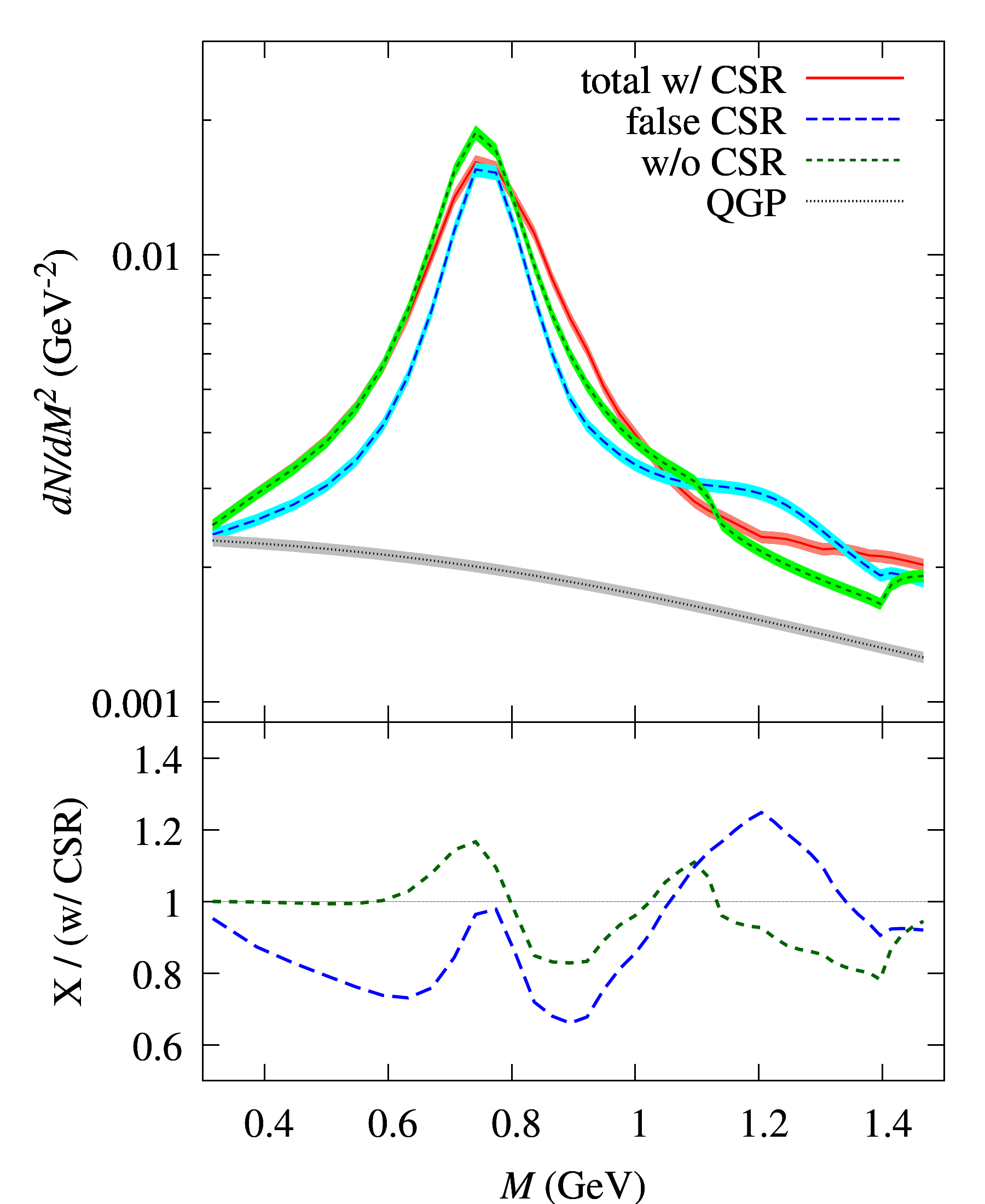}
\caption{(Color online).
Upper panel:
The total dilepton yields with the three hadronic spectral functions
for Pb+Pb collisions with collision energy
$\sqrt{s_\mathrm{NN}} = 2.76~\text{TeV}$
in the centrality 0--5\%.
The width of each line represents
statistical errors.
The dotted line is the rate from $\bar{q}q$ annihilation in the QGP medium.
Lower panel:
The production rates of the two scenarios, without CSR and false CSR, relative to the one with CSR.
}
\label{Fig:compare_s_total}
\end{figure}
Figure \ref{Fig:compare_s_total} shows the
total dilepton yield calculated with the three hadronic spectra via
Eq.~(\ref{Eq:total_rate}), integrated over the rapidity $\eta$ and transverse momenta $p_T$
in the ranges of $|\eta| < 0.8$ and $0.2 < p_T < 5.0$~GeV.
With the profile function~\cite{Shuryak:1978ij} evaluated under the current assumption (\ref{Eq:total_rate}),
the thermal dileptons produced in the hadronic phase amount to $70$\%.
The structures seen in Figs.~\ref{Fig:spectra_woCSR}-\ref{Fig:spectra_DEI} persist in the overall yields for all the three cases.
The consequences of CSR are observed above the $\rho$ peak: an increase appears compared with the case without CSR in $0.8 < M < 1$~GeV,
a window of the $a_1\rho\pi$ threshold effect operative, $1.1 < M < 1.4$~GeV, is filled in because of lighter $a_1$,
and those enhancements compensate a decrease at the $\rho$ peak.

The production rate in the false CSR scenario exhibits a striking contrast to the proper CSR.
The broadening effect is not captured at all. A substantial overestimate is found in $1.1 < M < 1.3$~GeV and becomes maximal at $M=1.2$~GeV,
amounting to a $25$\%-more contribution.
Clearly, the enhancement is a fake signature of CSR emerging at the wrong place in $M$. This is traced back to the spectral function based on
a naive superposition of the current correlators in matter-free-space, and with which the mass difference between $\rho$ and $a_1$ states,
$0.47$~GeV, is never resolved.

Confronted with the high-precision dimuon data by NA60 in Ref.~\cite{vanHees:2007th}, it was demonstrated that baryon effects via hadronic rescattering
in a hot medium are vital to reproduce the data from the low to intermediate $M$ region, whereas in $M > 0.8$~GeV there exists no such sensitivity.
Therefore, our observation based on the spectral functions that do not include apparent effects from baryons, is anticipated to stay unchanged
in a more realistic situation relevant to the ALICE experiments.

{\it Summary}
--
We studied the dilepton production
in hot QCD based on two distinct scenarios of chiral symmetry restoration (CSR), fully taking account of
space-time evolution of a created fireball in
relativistic heavy-ion collisions under the LHC conditions.
The scenario with CSR at which the degenerate $\rho$ and $a_1$
is realized, leading to {\it vanishing} chiral mixing, was compared with
the one in the false CSR scenario.
This has been often referred in the literature based on the low-temperature theorem, leading to a {\it maximal} chiral mixing.
We demonstrated that the two scenarios lead to distinct results as the signatures of CSR in dilepton production, especially in the most optimal window
of invariant mass, $1 < M < 1.4$~GeV.

The false CSR model causes a fake contribution being maximal, $25$\% more than in the proper CSR scenario, and this becomes evident at $M=1.24$~GeV, i.e.
the vacuum $a_1$ mass. The correct model describes an enhancement of the rate in the window of $1.1 < M < 1.4$~GeV, such that the in-medium $a_1$ fills in
this region separated by the onset of the $a_1\rho\pi$ threshold effect. The false CSR
scenario also results in a sizable underestimate below and above the $\rho$ peak
due to the lack of broadening via meson loops.

Our result illustrates that the widely-accepted conjecture, based on a naive application of low-temperature theorem toward $T_\chi$, should be interpreted
with caution. The appropriate description of chiral mixing produces a somewhat moderate but yet sizable contribution directly related with
the in-medium $a_1$, and the signal is supposed to survive in a more realistic situation at small net-baryon density in view of the detailed study
in a hadronic many-body approach~\cite{vanHees:2007th}.

In general, there exists a complicated admixture of two classes of chiral mixing in a hot and dense medium:
One is the mixing effect considered in this study, and another is induced solely by baryon density~\cite{Domokos:2007kt,Harada:2009cn}.
The latter never vanishes in striking contrast to the former, and results in a non-Breit-Wigner form of the vector spectral function
emerging near the CSR, thus serving as the measurable signature in dilepton production in cold and dense matter~\cite{Sasaki:2019jyh,Sasaki:2022vas}.
When a medium at moderate temperature and density is considered, it would be challenging to identify the effective chiral mixing implementing
the both classes. The systematic study as performed in this letter is restricted in a hot and dilute medium, relevant to the LHC physics, and the two mixing effects
can be separated from one another only at such a limiting situation.

{\it Acknowledgments}
--
The work was supported in part by the World Premier International Research Center
Initiative (WPI) under MEXT, Japan (CN, CS, KS, SY)
and by Japan Society for the Promotion of Science
(JSPS) KAKENHI Grant Nos.~JP20K03927, JP23H05439 (MH), JP17K05438, JP20H00156, JP20H11581 (CN), JP18H05401 and JP20H00163 (KS).
CS acknowledges the support by the Polish National
Science Centre (NCN) under OPUS Grant Nos.~2018/31/B/ST2/01663 and 2022/45/B/ST2/01527.
The numerical calculations were carried out on Yukawa-21 at YITP in Kyoto University, Japan.

\appendix
\bibliography{References}

\end{document}